\documentclass[aps,pre,reprint,superscriptaddress,nopacs,nofootinbib]{revtex4-1}

\usepackage{amsmath,amssymb,latexsym}
\usepackage{graphicx}
\graphicspath{{../fig/}}
\usepackage{hyperref}
\usepackage{txfonts}
\usepackage{mathrsfs}
\usepackage{color}
\DeclareSymbolFont{symbols}{OMS}{cmsy}{m}{n}
\DeclareSymbolFont{largesymbols}{OMX}{cmex}{m}{n}
\renewcommand{\bm}[1]{\boldsymbol #1}

\begin{document}

\title{
Out-of-Time-Order Fluctuation-Dissipation Theorem
}

\author{Naoto Tsuji}
\affiliation{RIKEN Center for Emergent Matter Science (CEMS), Wako 351-0198, Japan}
%\affiliation{PRESTO, Japan Science and Technology Agency, 4-1-8 Honcho Kawaguchi, Saitama 332-0012, Japan}
\author{Tomohiro Shitara}
\affiliation{Department of Physics, University of Tokyo, Hongo, Tokyo 113-0033, Japan}
\author{Masahito Ueda}
\affiliation{Department of Physics, University of Tokyo, Hongo, Tokyo 113-0033, Japan}
\affiliation{RIKEN Center for Emergent Matter Science (CEMS), Wako 351-0198, Japan}
%\email[]{}
%\homepage[]{}
%\thanks{}
%\altaffiliation{}

\begin{abstract}
We prove a generalized fluctuation-dissipation theorem for a certain class of
out-of-time-ordered correlators (OTOCs) with a modified statistical average, which we call bipartite OTOCs,
for general quantum systems in thermal equilibrium. The difference between the
bipartite and physical OTOCs defined by the usual statistical average is quantified by a measure of quantum fluctuations
known as the Wigner-Yanase skew information.
%which is a measure of quantum fluctuations.
Within this difference,
the theorem describes a universal relation 
between chaotic behavior in quantum systems
and a nonlinear-response function that involves a time-reversed process.
We show that the theorem can be generalized to higher-order $n$-partite OTOCs as well as
in the form of generalized covariance.
\end{abstract}

%\collaboration{}
%\noaffiliation

\date{\today}

\pacs{
%74.25.N-, 74.40.Gh, 71.10.Fd
%74: Superconductivity
%74.25.N-: Response to electromagnetic fields
%74.40.Gh: Nonequilibrium superconductivity
%71.10.Fd: Lattice fermion models (Hubbard model, etc.)
}

\maketitle

\section{Introduction}

The fluctuation-dissipation theorem (FDT) 
relates nonequilibrium transport coefficients to equilibrium fluctuations,
and plays a pivotal role in statistical mechanics.
%refers to a fundamental relation in statistical mechanics that bridges
%thermal fluctuation in equilibrium and response or relaxation against a perturbation near equilibrium. 
It dates back to Einstein's theory of Brownian motion \cite{Einstein1905} 
and the Nyquist relation between resistance and a thermal noise in voltage
\cite{Nyquist1928}, culminating in linear response theory \cite{Kubo1957}
(for a review, see, e.g., Ref.~\cite{Marconi2008}).
%The general fluctuation-dissipation theorem has been derived in accordance with
%the establishment of linear response theory \cite{Kubo1957}.

The FDT establishes the relationship between the expectation values of the commutator and the anticommutator,
\begin{align}
C_{[A,B]}(t,t')
&\equiv
\langle [\hat A(t),\hat B(t')]\rangle,
\\
C_{\{A,B\}}(t,t')
&\equiv
\langle \{\hat A(t),\hat B(t')\}\rangle,
\end{align}
of arbitrary (bosonic or fermionic\footnote{We call an operator $\hat A$ bosonic (fermionic) 
%if $\hat A$ is Grassmann even (odd).}})
if $\hat A$ is a linear combination of operators,
each of which contains an even (odd) number of fermion creation and/or annihilation operators.}) 
Heisenberg operators $\hat A(t)=e^{\frac{i}{\hbar}\hat H t}\hat A e^{-\frac{i}{\hbar}\hat H t}$
and $\hat B(t)=e^{\frac{i}{\hbar}\hat H t}\hat B e^{-\frac{i}{\hbar}\hat H t}$.
Here $\hat H$ is the Hamiltonian of the system, $\hbar$ is the Planck constant,
$\langle \cdot \rangle\equiv {\rm Tr}(\hat\rho\, \cdot \,)$,
and $\hat\rho=e^{-\beta\hat H}/Z$ ($Z={\rm Tr}e^{-\beta\hat H}$)
with $\beta=(k_B T)^{-1}$ being the inverse temperature ($k_B$ is the Boltzmann constant).
In the Fourier representation [i.e., $C_{\{A,B\}}(\omega)=\int_{-\infty}^{\infty} dt\, e^{i\omega t}C_{\{A,B\}}(t,0)$, etc.], 
the FDT is expressed as
\begin{align}
C_{\{A,B\}}(\omega)
&=
\coth\left(\frac{\beta\hbar\omega}{2}\right)
C_{[A,B]}(\omega).
\label{FDT}
\end{align}
If either $\hat A$ or $\hat B$ is bosonic, then
$C_{\{A,B\}}(\omega)$ represents thermal fluctuations and $C_{[A,B]}(\omega)$ represents
dissipation (and vice versa if both $\hat A$ and $\hat B$ are fermionic) \cite{Kubo1957,KuboBook,noneqDMFTreview}.

%The FDT is commonly expressed as
%\begin{align}
%C_{\{A,B\}}(\omega)
%&=
%\coth\left(\frac{\beta\hbar\omega}{2}\right)
%C_{[A,B]}(\omega),
%\label{FDT}
%\end{align}
%where $\hat A$ and $\hat B$ are arbitrary linear operators (irrespective of bosonic or fermionic),
%$\beta=(k_B T)^{-1}$ is the inverse temperature, 
%and $C_{\{A,B\}}(\omega)$ and $C_{[A,B]}(\omega)$ are 
%the Fourier transformed functions of
%\begin{align}
%C_{\{A,B\}}(t,t')
%&\equiv
%\langle \{\hat A(t),\hat B(t')\}\rangle,
%\\
%C_{[A,B]}(t,t')
%&\equiv
%\langle [\hat A(t),\hat B(t')]\rangle,
%\end{align}
%[i.e., $C_{\{A,B\}}(\omega)=\int_{-\infty}^{\infty} dt\, e^{i\omega t}C_{\{A,B\}}(t,0)$, etc.].
%Here $\langle \cdot \rangle\equiv {\rm Tr}(\hat\rho\, \cdot \,)$ is the expectation value,
%%(i.e., $C_{\{A,B\}}(\omega)=\int dt e^{i\omega t}C_{\{A,B\}}(t,0)$), 
%$\hat\rho=e^{-\beta\hat H}/Z$ ($Z={\rm Tr}e^{-\beta\hat H}$) is the thermal density matrix,
%$\hat H$ is the system Hamiltonian,
%and the time-dependent operators are defined in the Heisenberg picture
%(i.e., $\hat A(t)=e^{\frac{i}{\hbar}\hat H t}\hat A e^{-\frac{i}{\hbar}\hat H t}$).
%The relation (\ref{FDT}) holds for arbitrary operators $\hat A$ and $\hat B$ (irrespective of bosonic or fermionic). 
%When either $\hat A$ or $\hat B$ is bosonic,
%$C_{\{A,B\}}(\omega)$ represents the thermal fluctuation, while $C_{[A,B]}(\omega)$ represents
%the dissipation (and vice versa for both $\hat A$ and $\hat B$ being fermionic).

What is the law that governs higher-order fluctuations beyond the FDT (\ref{FDT}) and beyond the linear response regime?
The generalization of the FDT has led to deeper understanding of nonequilibrium statistical mechanics. 
The prime examples are the fluctuation theorem \cite{EvansCohenMorriss1993,EvansSearles1994} and the Jarzynski equality \cite{Jarzynski1997}, 
which are valid in arbitrary far off-equilibrium situations,
reproduce the FDT (\ref{FDT}) at zero frequency if applied to near thermal equilibrium, and place constraints
on higher-order fluctuations \cite{Gallavotti1996,AndrieuxGaspard2007,SaitoUtsumi2008,EspositoHarbolaMukamel2009,Campisi2011}.

Here we pursue a different direction of generalization of the FDT by
considering the second moments of fluctuation and dissipation such as 
$\langle \{\hat A(t),\hat B(t')\}^2\rangle$ and $\langle [\hat A(t),\hat B(t')]^2\rangle$.
They involve the operator sequences $\hat A(t)\hat B(t')\hat A(t)\hat B(t')$ and $\hat B(t')\hat A(t)\hat B(t')\hat A(t)$
that constitute out-of-time-ordered correlators (OTOCs) \cite{LarkinOvchinnikov1969}.

The OTOC has attracted growing attention as a measure to characterize chaotic behavior in quantum systems \cite{Kitaev2015}.
The relation to chaos can be seen in the semiclassical approximation:
If $\hat A$ and $\hat B$ form a canonically conjugate pair, then $\langle [\hat A(t),\hat B(0)]^2\rangle
\sim -\hbar^2 \langle\!\langle \{A(t),B(0)\}_P^2\rangle\!\rangle
=-\hbar^2\langle\!\langle \big(\frac{\partial A(t)}{\partial A(0)}\big)^2 \rangle\!\rangle$,
where $\langle\!\langle \cdot \rangle\!\rangle$ is the classical phase-space average with respect to the Gibbs ensemble,
and $\{,\}_P$ is the Poisson bracket. This quantity indicates the sensitivity of the time-evolving quantity $A(t)$
to its initial value $A(0)$ and is expected to grow exponentially in time for chaotic systems (``butterfly effect'')
as $\sim e^{\lambda t}$, where $\lambda$ is an analog of the Lyapunov exponent in classical chaotic systems
(see also Ref.~\cite{Rozenbaum2016}).
The interest in OTOCs has recently surged in various contexts
including the Sachdev-Ye-Kitaev model \cite{SachdevYe1993,Kitaev2015,MaldacenaStanford2016},
black holes and the holography principle \cite{ShenkerStanford2014b,ShenkerStanford2015,MaldacenaShenkerStanford2016}, 
quantum information \cite{Hosur2016,Swingle2016,Campisi2016},
many-body localization \cite{Huang2016,Fan2016,Chen2016,He2016}, and
strongly correlated systems \cite{Shen2016,Aleiner2016,TsujiWernerUeda2016,Dora2016,Bohrdt2016}.
The OTOC has recently been observed in experiments \cite{Garttner2017,Li2016,Wei2016,Meier2017}.

In this paper, we show that a generalized fluctuation-dissipation theorem holds for a certain class of OTOCs
with an arbitrary frequency.
The theorem describes a universal relation between chaotic properties in quantum systems and a nonlinear response function
for a perturbation involving a time-reversed process. To be more precise, there is a difference in operator ordering between
OTOCs defined by the usual statistical average [$\langle \hat A(t)\hat B(t')\hat A(t)\hat B(t')\rangle
={\rm Tr}(\hat\rho \hat A(t)\hat B(t')\hat A(t)\hat B(t'))$] and those that do obey the out-of-time-order FDT.
This difference can be expressed in terms of the Wigner-Yanase skew information \cite{WignerYanase1963}
which is known in the context of quantum information theory and serves as a measure of information contents contained in quantum fluctuations
of observables.
Within the difference of the skew information, the out-of-time-order FDT relates the chaotic behavior and the nonlinear
response function.

The rest of this paper is organized as follows. In Sec.~\ref{main results}, 
we present the statement of one of the main results in the paper, the out-of-time-order FDT.
In Sec.~\ref{physical meaning}, we discuss the physical meaning of the out-of-time-order FDT.
We prove the out-of-time-order FDT in Sec.~\ref{proof}.
In Sec.~\ref{generalization}, we generalize the theorem to higher-order OTOCs as well as other operator ordering of OTOCs. 
In Sec.~\ref{conclusion}, we conclude the paper. 
In Appendix, we present the proofs of some relations among OTOCs used in the main text.

\section{Main results}
\label{main results}

The FDT is generalized for OTOCs not in a straightforward manner
but in a twisted form. Namely,
we should split $\hat\rho$ into two $\hat\rho^{\frac{1}{2}}$'s, one of which is inserted 
in between commutators and/or anticommutators
of $\hat A(t)$ and $\hat B(t')$ and the other is placed in front of them.
To be specific, we define a {\it bipartite} OTOC (also called a regularized OTOC)
\cite{MaldacenaShenkerStanford2016,Stanford2016,Yao2016,PatelSachdev2016} as
\begin{align}
C_{AB}^{\alpha_1\alpha_2}(t,t')
&\equiv
C_{[A,B]_{\alpha_1}[A,B]_{\alpha_2}}(t,t')
\notag
\\
&\equiv
{\rm Tr}\left(
\hat\rho^{\frac{1}{2}}[\hat A(t),\hat B(t')]_{\alpha_1}
\hat\rho^{\frac{1}{2}}[\hat A(t),\hat B(t')]_{\alpha_2}
\right),
\label{bipartite OTOC}
\end{align}
where $\alpha_1, \alpha_2=\pm$, and $[, ]_{-(+)}$ represents the (anti) commutator.
%They appear in Refs.~\cite{MaldacenaShenkerStanford2016,Stanford2016,PatelSachdev2016}.
%Note the difference between (\ref{bipartite OTOC}) and a {\it physical} OTOC,
Note that (\ref{bipartite OTOC}) is different from an ordinary OTOC which takes the form of
the expectation value [${\rm Tr}(\hat\rho\cdots)$] of products of (anti)commutators for a given state $\hat\rho$,
\begin{align}
C_{AB}^{{\rm phys}, \alpha_1\alpha_2}(t,t')
&\equiv
C_{[A,B]_{\alpha_1}[A,B]_{\alpha_2}}^{\rm phys}(t,t')
\notag
\\
&\equiv
{\rm Tr}\left(
\hat\rho[\hat A(t),\hat B(t')]_{\alpha_1}
[\hat A(t),\hat B(t')]_{\alpha_2}
\right).
\label{physical OTOC}
\end{align}
Since this quantity is written in the form of the expectation value
that allows for a direct physical interpretation,
we shall refer to (\ref{physical OTOC}) as a {\it physical} OTOC.
%which has a direct physical meaning of the expectation value of products of (anti)commutators.
Depending on $\alpha_1,\alpha_2=\pm$, Eq.~(\ref{bipartite OTOC}) introduces four types of bipartite OTOCs,
of which $C_{\{A,B\}[A,B]}$ and $C_{[A,B]\{A,B\}}$ are equal due to the cyclic invariance of the trace.
Hence there are three independent bipartite OTOCs for a given pair of $\hat A$ and $\hat B$.

One of the main results in this paper is that for {\it any} quantum system in thermal equilibrium
the three bipartite OTOCs are related via
\begin{align}
C_{\{A,B\}^2}(\omega)+C_{[A,B]^2}(\omega)
&=
2\coth\left(\frac{\beta\hbar\omega}{4}\right)C_{\{A,B\}[A,B]}(\omega),
\label{FDT2}
\end{align}
which we call the out-of-time-order FDT.
If we ignore the difference in operator ordering between (\ref{bipartite OTOC}) and (\ref{physical OTOC})
%(which is justified in the classical limit),
(the physical meaning of this is explained in Sec.~\ref{physical meaning}),
then the equality (\ref{FDT2}) implies a universal relation among the second moments of fluctuation and dissipation, and their cross-correlation.
In this sense, the equality (\ref{FDT2}) can be viewed as a second-order extension of the FDT (\ref{FDT}).
%We will later show that the difference between the physical and bipartite OTOCs characterizes quantum fluctuations.

\section{Physical meaning of the out-of-time-order fluctuation-dissipation theorem}
\label{physical meaning}

To see the physical meaning of the equality (\ref{FDT2}),
let us first note that the difference between $C_{AB}^{\alpha_1\alpha_2}(t,t')$ (\ref{bipartite OTOC})
and $C_{AB}^{{\rm phys},\alpha_1\alpha_2}(t,t')$ (\ref{physical OTOC})
takes a form reminiscent of
the Wigner-Yanase (WY) skew information\footnote{There is a one-parameter generalization of the WY skew information due to Dyson, i.e., 
  $I_\alpha(\hat\rho,\hat O)={\rm Tr}(\hat\rho\hat O^2)-{\rm Tr}(\hat\rho^\alpha\hat O \rho^{1-\alpha}\hat O)$ ($0\leqslant \alpha \leqslant 1$).} \cite{WignerYanase1963} defined by
\begin{align}
I_{\frac{1}{2}}(\hat\rho,\hat O)
&\equiv
-\frac{1}{2}{\rm Tr}([\hat\rho^{\frac{1}{2}},\hat O]^2)
\notag
\\
&=
{\rm Tr}(\hat\rho \hat O^2)
-{\rm Tr}(\hat\rho^{\frac{1}{2}} \hat O \hat\rho^{\frac{1}{2}}\hat O)
\end{align}
for a Hermitian operator $\hat O$.
It serves as a measure of information contents concerning quantum fluctuations.
Here by quantum fluctuations we mean the following \cite{Luo2005}.
Let us consider the variance of $\hat O$, $\langle (\Delta\hat O)^2\rangle$,
where $\Delta\hat O\equiv\hat O-\langle\hat O\rangle$. 
The variance $\langle (\Delta\hat O)^2\rangle$ generally contains
classical mixing and quantum uncertainty, so that
we are tempted to decompose the variance
as 
\begin{align}
\langle (\Delta\hat O)^2\rangle
&=
C(\hat\rho,\hat O)+Q(\hat\rho,\hat O). 
\end{align}
If $C(\hat\rho,\hat O)$ and $Q(\hat\rho,\hat O)$
satisfy the following conditions, we call them the classical and quantum fluctuations of $\hat O$:
\begin{enumerate}
\renewcommand{\labelenumi}{(\alph{enumi})}
\item $C(\hat\rho,\hat O), Q(\hat\rho,\hat O)\geqslant 0$.
\item If $\hat\rho$ is pure, then $C(\hat\rho,\hat O)=0$ and $Q(\hat\rho,O)=\langle(\Delta\hat O)^2\rangle$.
\item If $\hat\rho$ and $\hat O$ commute, then $C(\hat\rho,O)=\langle(\Delta\hat O)^2\rangle$ and 
$Q(\hat\rho,\hat O)=0$. 
\item $C(\hat\rho,\hat O)$ is concave and $Q(\hat\rho,O)$ is convex as functions of $\hat\rho$,
i.e., 
\begin{align}
C(\lambda\hat\rho_1+(1-\lambda)\hat\rho_2,\hat O)\geqslant \lambda C(\hat\rho_1,\hat O)+(1-\lambda)C(\hat\rho_2,\hat O),
\notag
\\
Q(\lambda\hat\rho_1+(1-\lambda)\hat\rho_2,\hat O)\leqslant \lambda Q(\hat\rho_1,\hat O)+(1-\lambda)Q(\hat\rho_2,\hat O),
\notag
\end{align}
for $0\leqslant\lambda\leqslant 1$.
\end{enumerate}
The condition (d) means that classical fluctuations should increase and quantum fluctuations should decrease by a classical mixing of states. 
These conditions are in accordance with our intuition of quantum fluctuations.

Although such a decomposition is not unique \cite{Hansen2008}, the WY skew information provides one realization 
of the measure of quantum fluctuations [$Q(\hat\rho,\hat O)=I_{\frac{1}{2}}(\hat\rho,\hat O)$]. In fact, it satisfies the inequalities
\begin{align}
0\leqslant I_{\frac{1}{2}}(\hat\rho,\hat O)
\leqslant \langle (\Delta\hat O)^2\rangle.
\label{WYD inequality}
\end{align}
%where $\Delta\hat O\equiv\hat O-\langle\hat O\rangle$ and $\langle (\Delta\hat O)^2\rangle$
%represents the variance for an observable $\hat O$.
%\cite{variance}.
The equality on the left-hand side
of (\ref{WYD inequality}) is satisfied when $[\hat\rho,\hat O]=0$,
and the one on the right-hand side is met when $\hat\rho$ is
a pure state. 
%\textcolor{red}{
%For an extensive observable $\hat O$, $\langle(\Delta\hat O)^2\rangle$ usually behaves as $O(N)$ ($N$ is the number of particles),
%which means that $I_{\frac{1}{2}}(\hat\rho,\hat O)$ is much smaller than ${\rm Tr}(\hat\rho\hat O^2)\sim {\rm Tr}(\hat\rho^{\frac{1}{2}}\hat O\hat\rho^{\frac{1}{2}}\hat O)\sim O(N^2)$.
%}
Furthermore, the WY skew information is convex as a function of a quantum state \cite{Lieb1973},
\begin{align}
I_{\frac{1}{2}}(\lambda\hat\rho_1+(1-\lambda)\hat\rho_2,\hat O)
\leqslant
\lambda I_{\frac{1}{2}}(\hat\rho_1,\hat O)+(1-\lambda)I_{\frac{1}{2}}(\hat\rho_2,\hat O),
\label{convexity}
\end{align}
for $0\leqslant \lambda \leqslant 1$. That is,
%The inequality (\ref{convexity}) represents that the WY skew information 
it decreases under a classical mixing of quantum states,
justifying [with (\ref{WYD inequality})] the use of $I_{\frac{1}{2}}(\hat\rho,\hat O)$ as an information-theoretic measure of quantum fluctuations. 

%The WYD skew information also satisfies the convexity inequality \cite{Lieb1973},
%\begin{align}
%I_\alpha(\lambda\hat\rho_1+(1-\lambda)\hat\rho_2,\hat O)
%\le
%\lambda I_\alpha(\hat\rho_1,\hat O)+(1-\lambda)I_\alpha(\hat\rho_2,\hat O),
%\end{align}
%which justifies [with (\ref{WYD inequality})] the use of $I_\alpha(\hat\rho,\hat O)$ as an information measure. 

If $\hat A$ and $\hat B$ are Hermitian, then
$C_{AB}^{\alpha_1\alpha_2}(t,t')$ and $C_{AB}^{{\rm phys},\alpha_1\alpha_2}(t,t')$ are related 
to the WY skew information via
\begin{align}
C_{\{A,B\}^2}(t,t')
&=
C_{\{A,B\}^2}^{\rm phys}(t,t')-I_{\frac{1}{2}}\!\left(\hat\rho, \{\hat A(t),\hat B(t')\}\right),
\label{difference1}
\\
C_{[A,B]^2}(t,t')
&=
C_{[A,B]^2}^{\rm phys}(t,t')+I_{\frac{1}{2}}\!\left(\hat\rho, i[\hat A(t),\hat B(t')]\right),
\label{difference2}
\\
C_{\{A,B\}[A,B]}(t,t')
&=
\frac{1}{2}[C_{\{A,B\}[A,B]}^{\rm phys}(t,t')+C_{[A,B]\{A,B\}}^{\rm phys}(t,t')]
\notag
\\
&\quad
+\frac{i}{4}I_{\frac{1}{2}}\!\left(\hat\rho,\{\hat A(t),\hat B(t')\}+i[\hat A(t),\hat B(t')]\right)
\notag
\\
&\quad
-\frac{i}{4}I_{\frac{1}{2}}\!\left(\hat\rho,\{\hat A(t),\hat B(t')\}-i[\hat A(t),\hat B(t')]\right).
\label{difference3}
\end{align}
Note that $\{\hat A(t),\hat B(t')\}$, $i[\hat A(t),\hat B(t')]$, and $\{\hat A(t),\hat B(t')\}\pm i[\hat A(t),\hat B(t')]$ are Hermitian.
%(while $\hat A$ and $\hat B$ themselves are not necessarily hermitian).
We thus find that the difference between the bipartite (\ref{bipartite OTOC}) and physical OTOCs (\ref{physical OTOC})
can be expressed in terms of the skew information.
%\textcolor{red}{which has not been known before.}
Within this difference, which is negligible when quantum fluctuations are small,
Eq.~(\ref{FDT2}) shows the relation among 
the second moments of fluctuation and dissipation, and their cross-correlation.
We can explicitly express this by rewriting Eq.~(\ref{FDT2}) in terms of the physical OTOCs,
\begin{align}
&
C_{\{A,B\}^2}^{\rm phys}(\omega)+C_{[A,B]^2}^{\rm phys}(\omega)
\notag
\\
&=
\coth\left(\frac{\beta\hbar\omega}{4}\right)
[C_{\{A,B\}[A,B]}^{\rm phys}(\omega)+C_{[A,B]\{A,B\}}^{\rm phys}(\omega)]
+I_{AB}(\omega),
\label{FDT2 physical}
\end{align}
where
\begin{align}
I_{AB}(\omega)
&=
\int_{-\infty}^{\infty} dt e^{i\omega t}\bigg\{
I_{\frac{1}{2}}\!\left(\hat\rho, \{\hat A(t),\hat B(0)\}\right)
-I_{\frac{1}{2}}\!\left(\hat\rho, i[\hat A(t),\hat B(0)]\right)
\notag
\\
&\quad
+\frac{i}{2}\coth\left(\frac{\beta\hbar\omega}{4}\right)
I_{\frac{1}{2}}\!\left(\hat\rho,\{\hat A(t),\hat B(0)\}+i[\hat A(t),\hat B(0)]\right)
\notag
\\
&\quad
-\frac{i}{2}\coth\left(\frac{\beta\hbar\omega}{4}\right)
I_{\frac{1}{2}}\!\left(\hat\rho,\{\hat A(t),\hat B(0)\}-i[\hat A(t),\hat B(0)]\right)
\bigg\}
\end{align}
is a linear combination of the skew information.

\begin{figure}[t]
\includegraphics[width=8.5cm]{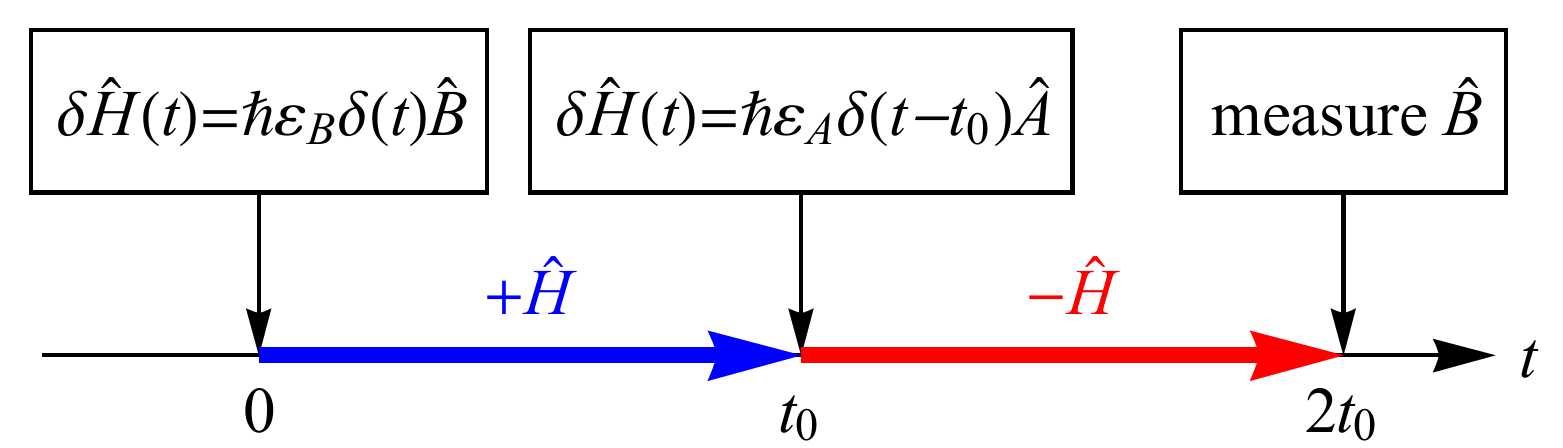}
\caption{Measurement protocol for the nonlinear response function $L_{(AB)^2}^{(3)}(t_0,0)$.
The system is perturbed by the pulsed fields at $t=0$ and $t=t_0$. The system evolves with the Hamiltonian $+\hat H$
from $t=0$ to $t=t_0$ and then with $-\hat H$ from $t=t_0$ to $t=2t_0$. The measurement of $\hat B$ is performed
at $t=2t_0$.}
\label{protocol}
\end{figure}

The physical meaning of Eq.~(\ref{FDT2}) [or Eq.~(\ref{FDT2 physical})] is as follows. 
The right-hand side of (\ref{FDT2}) is related to a certain type of a nonlinear-response function (Fig.~\ref{protocol}).
To see this, let us consider the following experimental protocol.
The initial state is set to be in thermal equilibrium with $\hat\rho$.
At time $t=0$, we perturb the system with a variation of the Hamiltonian $\delta \hat H(t)=\hbar\varepsilon_B \delta(t)\hat B$ 
($\varepsilon_B\in\mathbb R$).
Then we let the system evolve from $t=0$ to $t_0$ with the Hamiltonian $+\hat H$. At $t=t_0$ $(>0)$, 
we perturb the system with
$\delta\hat H(t)=\hbar\varepsilon_A \delta(t-t_0)\hat A$ ($\varepsilon_A\in\mathbb R$). Then we let the system
evolve from $t=t_0$ to $2t_0$ with the inverted Hamiltonian $-\hat H$ 
(as in spin echo, Loschmidt echo, or ultracold atom \cite{TsujiWernerUeda2016} experiments),
%(we assume to have a full control of the whole quantum system such as ultracold-atom systems \cite{TsujiWernerUeda2016}
%so that $\hat H$ can be inverted with sufficient accuracy), 
i.e., the time propagation is effectively reversed. 
Finally, we measure $\hat B$ at $t=2t_0$.
Let us suppose that $\varepsilon_A$
and $\varepsilon_B$ are sufficiently small,
which allows us to expand $\langle \hat B(2t_0)\rangle$ with respect to $\varepsilon_A$ and $\varepsilon_B$,
\begin{align}
\langle \hat B(2t_0)\rangle
&=
\sum_{m,n=0}^\infty \varepsilon_A^m \varepsilon_B^n
[\delta_{A^mB^n}^{m+n}\langle \hat B(2t_0)\rangle].
\end{align}
Here $\delta_{A^mB^n}^{m+n}\langle \hat B(2t_0)\rangle$ represents the expansion coefficient
at the $m$th and $n$th orders with respect to the perturbation strength $\varepsilon_A$ and $\varepsilon_B$.
The lowest order at which OTOCs appear is the third order ($m+n=3$).
We define a nonlinear response function $L_{(AB)^2}^{(3)}(t_0,0)$
as a coefficient of the third-order variation of $\langle \hat B(2t_0)\rangle$
that is proportional to $\varepsilon_A^2 \varepsilon_B$, i.e., 
\begin{align}
\delta_{A^2B}^3\langle \hat B(2t_0)\rangle
&=:
\frac{1}{2}L_{(AB)^2}^{(3)}(t_0,0).
\label{B(2t)}
\end{align}
The density matrix at $t=2t_0$ is given by
\begin{align}
\hat\rho(2t_0)
&=
e^{\frac{i}{\hbar}\hat Ht_0}e^{-i\varepsilon_A\hat A}e^{-\frac{i}{\hbar}\hat Ht_0}e^{-i\varepsilon_B\hat B}
\hat\rho e^{i\varepsilon_B\hat B}e^{\frac{i}{\hbar}\hat Ht_0}e^{i\varepsilon_A\hat A}e^{-\frac{i}{\hbar}\hat Ht_0}.
\end{align}
By expanding $\langle \hat B(2t_0)\rangle={\rm Tr}[\hat\rho(2t_0)\hat B]$
with respect to $\varepsilon_A$ and $\varepsilon_B$, and collecting terms proportional to $\varepsilon_A^2\varepsilon_B$,
we obtain
\begin{align}
L_{(AB)^2}^{(3)}(t_0,0)
&=
-i[C_{\{A,B\}[A,B]}^{\rm phys}(t_0,0)+C_{[A,B]\{A,B\}}^{\rm phys}(t_0,0)]
\notag
\\
&\quad
+iC_{[A^2,B^2]}(t_0,0).
\label{L_(AB)^2}
\end{align}
The two terms in the square bracket in Eq.~(\ref{L_(AB)^2}) coincide with those in
the square bracket in Eq.~(\ref{difference3}) (with $t=t_0$ and $t'=0$)\footnote{The present protocol measures $L_{(AB)^2}^{(3)}(t_0,0)$ for $t_0>0$. The remaining part ($t_0<0$) 
  can be obtained by changing the protocol such that the system evolves with $-H$ from $t=0$ to $t_0$ and with $+H$ from $t=t_0$ to $2t_0$.},
while the last term in Eq.~(\ref{L_(AB)^2}), $L_{A^2B^2}^{(1)}(t_0,0)\equiv -iC_{[A^2,B^2]}(t_0,0)$,
is a linear-response function with respect to $\hat A^2$ and $\hat B^2$ 
(i.e., a linear response of $\langle \hat A(t_0)^2\rangle$
against a perturbation $\delta\hat H(t)=\hbar\varepsilon_B\delta(t)\hat B^2$),
which can be measured independently.
One can see that the right-hand side of (\ref{FDT2}), $2C_{\{A,B\}[A,B]}(t,t')$, is related to the response function
$i[L_{(AB)^2}^{(3)}(t,t')+L_{A^2B^2}^{(1)}(t,t')]$
within the difference of the WY skew information,
\begin{align}
2C_{\{A,B\}[A,B]}(t,t')
&=
i[L_{(AB)^2}^{(3)}(t,t')+L_{A^2B^2}^{(1)}(t,t')]
\notag
\\
&\quad
+\frac{i}{4}I_{\frac{1}{2}}\!\left(\hat\rho,\{\hat A(t),\hat B(t')\}+i[\hat A(t),\hat B(t')]\right)
\notag
\\
&\quad
-\frac{i}{4}I_{\frac{1}{2}}\!\left(\hat\rho,\{\hat A(t),\hat B(t')\}-i[\hat A(t),\hat B(t')]\right).
\label{RHS FDT2}
\end{align}
%via Eqs.~(\ref{difference3}) and (\ref{L_(AB)^2}).
%up to the difference of the skew information and the linear response function.

The protocol has advantages
that it can be applied to arbitrary thermal initial states and avoids multiple measurements that cause 
measurement back action.
This is in contrast to the protocols described in Refs.~\cite{TsujiWernerUeda2016,Garttner2017},
where the initial state is set to be an eigenstate of the operator $\hat A$ or $\hat B$ to readout the OTOC.
This is equivalent to making a projection measurement at the initial time, which causes measurement back actions.
A Loschmidt-echo-type protocol similar to the one shown in Fig.~\ref{protocol} has been proposed in Ref.~\cite{Swingle2016}.
The difference is that in the former one measures the nonlinear response function $L_{(AB)^2}(t,0)$
to reconstruct ${\rm Re}\langle \hat A(t)\hat B(0)\hat A(t)\hat B(0)\rangle
=\frac{1}{4}[C_{\{A,B\}^2}^{\rm phys}(t,0)+C_{[A,B]^2}^{\rm phys}(t,0)]$ 
for Hermitian operators $\hat A$ and $\hat B$
via the out-of-time-order FDT (\ref{FDT2}), 
while in the latter one measures 
$|\langle\psi|\hat W^\dagger(t)\hat V^\dagger(0)\hat W(t)\hat V(0)|\psi\rangle|^2$ for unitary operators $\hat V$ and $\hat W$.
The latter also requires the projection onto the initial state $|\psi\rangle$.
We note that there are various other types of protocols which have been proposed
to measure OTOCs \cite{Swingle2016,Yao2016,Zhu2016,Bohrdt2016,YungerHalpern2017,YungerHalpernSwingleDressel2017}.
%needs a coupling to ancilla systems, while our protocol does not.

The left-hand side of (\ref{FDT2}), on the other hand, is related to chaotic behavior in quantum many-body systems \cite{Kitaev2015,MaldacenaShenkerStanford2016}. 
As we have mentioned, if $\hat A$ and $\hat B$ are a canonically conjugate pair, then
$C_{[A,B]^2}^{\rm phys}(t,0)\sim -\hbar^2 \langle\!\langle \big(\frac{\partial A(t)}{\partial A(0)}\big)^2\rangle\!\rangle$
in the semiclassical regime, indicating an initial-value sensitivity of $A(t)$.
In chaotic systems, $-\hbar^2 \langle\!\langle \big(\frac{\partial A(t)}{\partial A(0)}\big)^2\rangle\!\rangle$
is expected to grow exponentially in time ($\sim e^{\lambda t}$), 
where $\lambda$ is an analog of the Lyapunov exponent.
%The characteristic behavior in chaotic systems is that this quantity grows exponentially in time ($\sim e^{\lambda t}$
%with $\lambda$ analogous to the Lyapunov exponent in classical chaotic systems; see also \cite{Rozenbaum2016}).
The exponential growth in $C_{[A,B]^2}^{\rm phys}(t,0)$ arises from
its out-of-time-ordered part $\langle (\hat A(t)\hat B(0))^2\rangle+\langle (\hat B(0)\hat A(t))^2\rangle$
\cite{MaldacenaShenkerStanford2016},
which is equal to $\frac{1}{2}[C_{\{A,B\}^2}^{\rm phys}(t,0)+C_{[A,B]^2}^{\rm phys}(t,0)]$.
Therefore, the left-hand side of Eq.~(\ref{FDT2}) represents an initial-value sensitivity of a time-evolving observable
(within the difference of the WY skew information).
Based on these observations, we are led to a general principle that the nonlinear response defined in Eq.~(\ref{B(2t)}) is related
to chaotic behavior in quantum systems through the out-of-time-order FDT (\ref{FDT2}).
This allows one to access the exponentially growing part of
the OTOC $\langle [\hat A(t),\hat B(0)]^2\rangle$ in chaotic systems by the nonlinear-response experiment.
As far as the exponential growth of $\langle [\hat A(t),\hat B(0)]^2\rangle$ is concerned,
the difference of the WY skew information, a measure of quantum fluctuations,
is suppressed in the semiclassical regime of our interest.

In the strictly classical limit with $\hbar\to 0$, 
%the quantum fluctuation becomes negligible, and 
%one can identify $C_{AB}^{\alpha_1\alpha_2}(t,t')$
% in (\ref{FDT2 Phi})
%with $C_{AB}^{{\rm phys},\alpha_1\alpha_2}(t,t')$.
%In this limit $\frac{1}{i\hbar}[\hat A(t),\hat B(t')]$ is replaced by 
%the Poisson bracket $\{A(t),B(t')\}_P$, 
%while $\{\hat A(t),\hat B(t')\}$ reduces to the product $2A(t)B(t')$.
%Therefore, the second term on the LHS of (\ref{FDT2}) vanishes for $\hbar\to 0$.
%and we obtain $C_{(AB)^2}(\omega)=\Phi_{AB}^2(\omega)$.
%From Eqs.~(\ref{FDT right}) and (\ref{Phi_AB^2 3}), the relation
%$i\hbar\partial_t\Phi_{AB}^2(t,t')=2k_BT C_{\{A,B\}[A,B]}(t,t')$ follows.
the out-of-time-order FDT (\ref{FDT2}) can be expressed as
\begin{align}
\partial_t \langle\!\langle A(t)^2 B(t')^2\rangle\!\rangle
&=
k_BT\langle\!\langle \{A(t)^2,B(t')^2\}_P\rangle\!\rangle.
\label{classical limit}
\end{align}
%where $\langle\!\langle \cdot \rangle\!\rangle$ denotes the classical phase-space average with respect to the Gibbs ensemble.
We can see that the classical limit of (\ref{FDT2}) reduces to that of the conventional FDT (\ref{FDT})
with $A(t)$ and $B(t)$ replaced by $A(t)^2$ and $B(t)^2$, respectively.
%brings about a nontrivial relation that involves the second moment of the fluctuation [i.e., the LHS of (\ref{classical limit})].
%\textcolor{red}{
%\sout{
%For chaotic systems in which $\langle [\hat A(t),\hat B(0)]^2\rangle$ grows exponentially in time
%(as in the case of the SYK model), 
%%\cite{Kitaev2015,MaldacenaStanford2016}
%the classical approximation
%may break down at the time scale of the order of the Ehrenfest time.
%%\cite{MaldacenaShenkerStanford2016}.
%Beyond this time scale, the relation (\ref{classical limit}) should be altered by (\ref{FDT2}).
%}
%}
%In this sense, the relation (\ref{classical limit}) can be used to diagnose the chaotic behavior
%in the classical regime.

\section{Proof of the out-of-time-order fluctuation-dissipation theorem}
\label{proof}

We now prove the equality (\ref{FDT2}). To this end, we introduce a representation of the bipartite OTOCs
different from (\ref{bipartite OTOC}):
\begin{align}
C_{AB}^{\mu_1\mu_2}(t,t')
&\equiv
{\rm Tr}\left(
\hat\rho^{\frac{1}{2}}(\hat A(t)\hat B(t'))^{\mu_1}
\hat\rho^{\frac{1}{2}}(\hat A(t)\hat B(t'))^{\mu_2}
\right),
\label{bipartite OTOC2}
\end{align}
where $\mu_1,\mu_2=\;>,<$, and
\begin{align}
(\hat A(t)\hat B(t'))^{\mu_i}
&\equiv
\begin{cases}
\hat A(t)\hat B(t') & \mbox{for} \quad \mu_i=\;>, \\
\hat B(t')\hat A(t) & \mbox{for} \quad \mu_i=\;<.
\end{cases}
\end{align}
In the above definition, we do not include the minus sign for $\mu_i=\;<$
when both $\hat A$ and $\hat B$ are fermionic.
However, all the arguments below can equally be applied to this case without any change.
Again we have $C_{AB}^{><}=C_{AB}^{<>}$
due to the cyclic invariance of the trace.
The two representations (\ref{bipartite OTOC}) and (\ref{bipartite OTOC2}) are connected 
by a linear transformation
\begin{align}
L
\begin{pmatrix}
C_{AB}^{>>} & C_{AB}^{><} \\
C_{AB}^{<>} & C_{AB}^{<<}
\end{pmatrix}
L^T
&=
\frac{1}{2}
\begin{pmatrix}
C_{[A,B]^2} & C_{[A,B]\{A,B\}} \\
C_{\{A,B\}[A,B]} & C_{\{A,B\}^2}
\end{pmatrix},
\label{matrix transformation}
\end{align}
where 
\begin{align}
L
&\equiv
\frac{1}{\sqrt{2}}
\begin{pmatrix}
1 & -1 \\
1 & 1
\end{pmatrix}
\end{align}
is an orthogonal matrix and $L^T$ is the transpose of $L$.
For convenience, we use notations $C_{(AB)^2}(t,t')\equiv C_{AB}^{>>}(t,t')$
and $C_{(BA)^2}(t',t)\equiv C_{AB}^{<<}(t,t')$.
We note the parallelism of the formulation with that for Keldysh Green's functions 
\cite{Keldysh1964,LarkinOvchinnikov1975,Rammer,Haehl2016}.

We first show that $C_{(AB)^2}$ and $C_{(BA)^2}$ are related to each other by
\begin{align}
C_{(BA)^2}(\omega)
&=
e^{\frac{\beta\hbar\omega}{2}}C_{(AB)^2}(-\omega).
\label{reciprocal relation}
\end{align}
This relation is analogous to the Kubo-Martin-Schwinger condition 
$C_{BA}(\omega)=e^{\beta\hbar\omega}C_{AB}(-\omega)$
\cite{Kubo1957,MartinSchwinger1959}
for conventional correlation functions 
$C_{AB}(t,t')\equiv{\rm Tr}(\hat\rho\hat A(t)\hat B(t'))$.
%and $C_{BA}(t,t')\equiv{\rm Tr}(\hat\rho\hat B(t')\hat A(t))$. 
The equality (\ref{reciprocal relation}) can be proven as follows.
We insert four complete sets of the eigenstates $\sum_k |k\rangle \langle k|$ of the Hamiltonian
$\hat H$ (with the eigenenergies $E_k$) in the definitions of $C_{(AB)^2}(t,t')$ and $C_{(BA)^2}(t,t')$,
obtaining
\begin{align}
C_{(AB)^2}(t,t')
&=
\frac{1}{Z}\sum_{k,l,m,n} e^{-\frac{\beta}{2}(E_k+E_m)}e^{\frac{i}{\hbar}(E_k-E_l+E_m-E_n)(t-t')}
\notag
\\
&\quad\times
\langle k|\hat A|l\rangle
\langle l|\hat B|m\rangle
\langle m|\hat A|n\rangle
\langle n|\hat B|k\rangle,
\\
C_{(BA)^2}(t,t')
&=
\frac{1}{Z}\sum_{k,l,m,n} e^{-\frac{\beta}{2}(E_k+E_m)}e^{\frac{i}{\hbar}(E_k-E_l+E_m-E_n)(t-t')}
\notag
\\
&\quad\times
\langle k|\hat B|l\rangle
\langle l|\hat A|m\rangle
\langle m|\hat B|n\rangle
\langle n|\hat A|k\rangle.
\label{C_(BA)^2 2}
\end{align}
By cyclically permuting the labels, $k\to l\to m\to n\to k$,
we have
\begin{align}
C_{(BA)^2}(t,t')
&=
\frac{1}{Z}\sum_{k,l,m,n} e^{-\frac{\beta}{2}(E_l+E_n)}e^{-\frac{i}{\hbar}(E_k-E_l+E_m-E_n)(t-t')}
\notag
\\
&\quad\times
\langle k|\hat A|l\rangle
\langle l|\hat B|m\rangle
\langle m|\hat A|n\rangle
\langle n|\hat B|k\rangle.
\end{align}
After the Fourier transformation, we obtain
\begin{align}
C_{(AB)^2}(\omega)
&=
\frac{1}{Z}\sum_{k,l,m,n} e^{-\frac{\beta}{2}(E_k+E_m)}
2\pi\delta\!\left(\omega+\tfrac{1}{\hbar}(E_k-E_l+E_m-E_n)\right)
\notag
\\
&\quad\times
\langle k|\hat A|l\rangle
\langle l|\hat B|m\rangle
\langle m|\hat A|n\rangle
\langle n|\hat B|k\rangle,
\label{C_(AB)^2 3}
\\
C_{(BA)^2}(\omega)
&=
\frac{1}{Z}\sum_{k,l,m,n} e^{-\frac{\beta}{2}(E_l+E_n)}
2\pi\delta\!\left(\omega-\tfrac{1}{\hbar}(E_k-E_l+E_m-E_n)\right)
\notag
\\
&\quad\times
\langle k|\hat A|l\rangle
\langle l|\hat B|m\rangle
\langle m|\hat A|n\rangle
\langle n|\hat B|k\rangle.
\label{C_(BA)^2 3}
\end{align}
Due to the presence of the $\delta$ function,
we can replace $E_l+E_n$ in the exponential in Eq.~(\ref{C_(BA)^2 3}) with 
$E_k+E_m-\hbar\omega$.
By comparing it with Eq.~(\ref{C_(AB)^2 3}), we obtain Eq.~(\ref{reciprocal relation}).

With the relation (\ref{reciprocal relation}),
the left-hand side of (\ref{FDT2}) is transformed as
\begin{align}
C_{\{A,B\}^2}(\omega)+C_{[A,B]^2}(\omega)
&=
2C_{(AB)^2}(\omega)+2C_{(BA)^2}(-\omega)
\notag
\\
&=
2(1+e^{-\frac{\beta\hbar\omega}{2}})C_{(AB)^2}(\omega),
\label{FDT left}
\end{align}
while the right-hand side is
\begin{align}
C_{\{A,B\}[A,B]}(\omega)
&=
C_{(AB)^2}(\omega)-C_{(BA)^2}(-\omega)
\notag
\\
&=
(1-e^{-\frac{\beta\hbar\omega}{2}})C_{(AB)^2}(\omega).
\label{FDT right}
\end{align}
Combining Eqs.~(\ref{FDT left}) and (\ref{FDT right}),
we arrive at the out-of-time-order FDT (\ref{FDT2}).
\hspace{\fill}$\blacksquare$

%To see how nontrivial the relation (\ref{FDT2}) is, we combine Eqs.~(\ref{FDT}) and (\ref{FDT2})
%and use 
%%the double-angle formula for the hyperbolic function, 
%$\coth\left(\beta\hbar\omega/2\right)=\frac{1}{2}\left[\coth\left(\beta\hbar\omega/4\right)
%+\coth\left(\beta\hbar\omega/4\right)^{-1}\right]$ to remove the temperature factor.
%We then obtain a nonlinear relation between usual correlation functions and OTOCs,
%\begin{align}
%\frac{C_{\{A,B\}}(\omega)}{C_{[A,B]}(\omega)}
%&=
%\frac{1}{2}\left[
%\frac{C_{\{A,B\}^2}(\omega)+C_{[A,B]^2}(\omega)}{2C_{\{A,B\}[A,B]}(\omega)}
%\right.
%\notag
%\\
%&\quad\left.
%+\frac{2C_{\{A,B\}[A,B]}(\omega)}{C_{\{A,B\}^2}(\omega)+C_{[A,B]^2}(\omega)}
%\right].
%\label{identity}
%\end{align}
%Equation (\ref{identity}) is an identity that holds for arbitrary linear operators $\hat A$ and $\hat B$ as long as
%the system is in thermal equilibrium.

Let us recall that the FDT (\ref{FDT}) can also be expressed as \cite{Kubo1957}
\begin{align}
%C_{\{A,B\}}(\omega)
%&=
%\beta\hbar\omega\coth\left(\frac{\beta\hbar\omega}{2}\right)\Phi_{AB}(\omega),
\beta\hbar\omega\,\Phi_{AB}(\omega)
&=
C_{[A,B]}(\omega),
\label{FDT Phi}
\end{align}
where $\Phi_{AB}$ is a canonical correlation (a quantum generalization of a classical correlator
$\langle\!\langle A(t)B(t')\rangle\!\rangle$),
\begin{align}
\Phi_{AB}(t,t')
&\equiv
\int_0^1 d\lambda {\rm Tr}\left(
\hat\rho^{1-\lambda}\hat A(t)\hat \rho^{\lambda}\hat B(t')
\right).
\label{Phi_AB}
\end{align}
Analogously to this,
the right-hand side of (\ref{FDT2}) can be rewritten in a form of a canonical bipartite OTOC
defined as
\begin{align}
\Phi_{(AB)^2}(t,t')
&\equiv
\int_0^1 d\lambda {\rm Tr}\left[\left(
\hat\rho^{\frac{1-\lambda}{2}}\hat A(t)\hat\rho^{\frac{\lambda}{2}}\hat B(t')
\right)^2\right].
\label{Phi_AB^2}
\end{align}
%We again insert the complete set of the eigenstates, obtaining
%\begin{align}
%&
%\Phi_{(AB)^2}(t,t')
%=
%\frac{1}{Z}\int_0^1 d\lambda
%\sum_{k,l,m,n} e^{-\frac{\beta}{2}(E_k+E_m)}
%e^{\frac{\lambda\beta}{2}(E_k-E_l+E_m-E_n)}
%\notag
%\\
%&\quad\times
%e^{\frac{i}{\hbar}(E_k-E_l+E_m-E_n)(t-t')}
%\langle k|\hat A|l\rangle
%\langle l|\hat B|m\rangle
%\langle m|\hat A|n\rangle
%\langle n|\hat B|k\rangle,
%\label{Phi_AB^2 2}
%\end{align}
%which is Fourier transformed into
%\begin{align}
%\Phi_{(AB)^2}(\omega)
%&=
%\frac{1}{Z}\int_0^1 d\lambda\, e^{-\frac{\lambda\beta}{2}\hbar\omega}
%\sum_{k,l,m,n} e^{-\frac{\beta}{2}(E_k+E_m)}
%\notag
%\\
%&\quad\times
%2\pi\delta\!\left(\omega+\tfrac{1}{\hbar}(E_k-E_l+E_m-E_n)\right)
%\notag
%\\
%&\quad\times
%\langle k|\hat A|l\rangle
%\langle l|\hat B|m\rangle
%\langle m|\hat A|n\rangle
%\langle n|\hat B|k\rangle
%\notag
%\\
%&=
%\frac{1-e^{-\frac{\beta\hbar\omega}{2}}}{\beta\hbar\omega/2}
%C_{(AB)^2}(\omega).
%\label{Phi_AB^2 3}
%\end{align}
The second-order extension of (\ref{FDT Phi}) is written as
\begin{align}
%C_{\{A,B\}^2}(\omega)+C_{[A,B]^2}(\omega)
%&=
%\beta\hbar\omega\coth\left(\frac{\beta\hbar\omega}{4}\right)\Phi_{(AB)^2}(\omega).
\beta\hbar\omega\,\Phi_{(AB)^2}(\omega)
&=
2C_{\{A,B\}[A,B]}(\omega).
\label{FDT2 Phi}
\end{align}
The proof of (\ref{FDT2 Phi}) is given as a special case of (\ref{GC FDT n Phi}) in Appendix \ref{appendix}.

To clarify the meaning of Eq.~(\ref{FDT2 Phi}), we note that
$\Phi_{(AB)^2}(t,t')$ can be written as
\begin{align}
\Phi_{(AB)^2}(t,t')
&=
\int_0^1 d\lambda\, C_{(AB)^2}\left(t-i\lambda\frac{\beta\hbar}{2},t'\right).
\label{regularization average}
\end{align}
That is, the time argument of the operator $\hat A$ is shifted to the direction of imaginary time.
This type of deformation has been employed to regularize OTOCs in the context of quantum field theory
\cite{MaldacenaShenkerStanford2016}.
Each $\lambda$ represents a different choice of regularization.
If the regularized OTOC $C_{(AB)^2}(t-i\lambda\frac{\beta\hbar}{2},0)$ shows an exponential growth
for every choice of regularization (in order for the growth to be physical, it should not depend on the choice of the regularization),
then its average (\ref{regularization average}) over the regularization parameter $\lambda$ also shows an exponential growth.
Then Eq.~(\ref{FDT2 Phi}) says that the averaged exponential growth of the OTOCs on the left-hand side
is related to the nonlinear response function on the right-hand side within the difference of the skew information 
[see Eq.~(\ref{RHS FDT2})].
%The relation (\ref{FDT2 Phi}) is useful for evaluating the RHS of (\ref{FDT2}) 
%if the calculation of different-time commutators is cumbersome.
%(note that there is an explicit $\hbar$ on the LHS of Eq.~(\ref{FDT2 Phi}) instead of commutators).
%$\Phi_{(AB)^2}(t,t')$ is explicitly calculated as
%\begin{align}
%&
%\Phi_{(AB)^2}(t,t')
%=
%\frac{1}{Z}
%\sum_{k,l,m,n} 
%\frac{e^{-\frac{\beta}{2}(E_l+E_n)}-e^{-\frac{\beta}{2}(E_k+E_m)}}{\frac{\beta}{2}(E_k-E_l+E_m-E_n)}
%\notag
%\\
%&\times
%e^{\frac{i}{\hbar}(E_k-E_l+E_m-E_n)(t-t')}
%\langle k|\hat A|l\rangle
%\langle l|\hat B|m\rangle
%\langle m|\hat A|n\rangle
%\langle n|\hat B|k\rangle.
%\label{Phi_AB^2 2}
%\end{align}
%From Eqs.~(\ref{FDT right}) and (\ref{Phi_AB^2 3}), we obtain
%If the matrix elements $\langle k|\hat A|l\rangle$ and $\langle l|\hat B|m\rangle$ are known,
%it is straightforward to evaluate $\Phi_{(AB)^2}(t,t')$
%and hence the RHS of Eq.~(\ref{FDT2}) via Eqs.~(\ref{FDT2 Phi}) and (\ref{Phi_AB^2 2}).

\section{Generalization of the out-of-time-order fluctuation-dissipation theorem}
\label{generalization}

The FDT (\ref{FDT2}) for OTOCs can be generalized in two ways.
One is to extend the relation to higher-order OTOCs \cite{Haehl2016,RobertsYoshida2017}.
%such as ${\rm Tr}(\hat\rho (\hat A(t)\hat B(t'))^n)$. Similarly to the second-order case,
%the FDT cannot be generalized in the physical form but in a modified form. 
Let us define an $n$-partite OTOC,
\begin{align}
C_{AB}^{\mu_1\mu_2\cdots\mu_n}(t,t')
&\equiv
{\rm Tr}\left[
\prod_{i=1}^n
\hat \rho^{\frac{1}{n}}
(\hat A(t)\hat B(t'))^{\mu_i}
\right],
\label{n-partite OTOC}
\end{align}
with $\mu_1,\mu_2,\dots,\mu_n=\;>,<$. In particular, we use abbreviations, $C_{(AB)^n}(t,t')\equiv C_{AB}^{>>\cdots >}(t,t')$
and $C_{(BA)^n}(t',t)\equiv C_{AB}^{<<\cdots <}(t,t')$. We perform a tensor transformation
\begin{align}
C_{AB}^{\alpha_1\alpha_2\cdots\alpha_n}(t,t')
&\equiv
2^{\frac{n}{2}}
\sum_{\mu_1,\dots,\mu_n}
{L^{\alpha_1}}_{\mu_1}
{L^{\alpha_2}}_{\mu_2}
\cdots
{L^{\alpha_n}}_{\mu_n}
C_{AB}^{\mu_1\mu_2\cdots\mu_n}(t,t')
\label{tensor transformation}
\end{align}
to switch to the commutator/anticommutator representation,
\begin{align}
C_{AB}^{\alpha_1\alpha_2\cdots\alpha_n}(t,t')
&=
C_{[A,B]_{\alpha_1}[A,B]_{\alpha_2}\cdots[A,B]_{\alpha_n}}(t,t')
\notag
\\
&=
{\rm Tr}\left(
\prod_{i=1}^n \hat\rho^{\frac{1}{n}}
[\hat A(t),\hat B(t')]_{\alpha_i}
\right),
\label{n-partite OTOC2}
\end{align}
with $\alpha_1,\alpha_2,\dots,\alpha_n=\pm$.
The transformation (\ref{tensor transformation}) is a higher-order generalization
of Eq.~(\ref{matrix transformation}). There are redundancies in the definitions (\ref{n-partite OTOC}) and (\ref{n-partite OTOC2}), 
$C_{AB}^{\mu_1\mu_2\cdots\mu_n}(t,t')=C_{AB}^{\mu_n\mu_1\cdots\mu_{n-1}}(t,t')$ and
$C_{AB}^{\alpha_1\alpha_2\cdots\alpha_n}(t,t')=C_{AB}^{\alpha_n\alpha_1\cdots\alpha_{n-1}}(t,t')$, 
due to the cyclic invariance of the trace.

In the same way as for $n=2$ [Eq.~(\ref{reciprocal relation})], we can prove (see Appendix \ref{appendix})
\begin{align}
C_{(BA)^n}(\omega)
&=
e^{\frac{\beta\hbar\omega}{n}}C_{(AB)^n}(-\omega)
\label{reciprocal relation n}
\end{align}
for arbitrary $n=1,2,3,\dots$. To rewrite the equality (\ref{reciprocal relation n}) in the form of the FDT,
we carry out (anti)symmetrization like Eqs.~(\ref{FDT left}) and (\ref{FDT right}),
\begin{align}
&
C_{(\{A,B\}+[A,B])^n}(\omega)\pm C_{(\{A,B\}-[A,B])^n}(\omega)
\notag
\\
&\qquad=
2^n[C_{(AB)^n}(\omega)\pm C_{(BA)^n}(-\omega)]
\notag
\\
&\qquad=
2^n(1\pm e^{-\frac{\beta\hbar\omega}{n}})C_{(AB)^n}(\omega).
\label{symmetrized n}
\end{align}
By taking the ratio of both sides of Eq.~(\ref{symmetrized n}) between the ones with $+$ and $-$ signs
and explicitly expanding $(\{A,B\}\pm [A,B])^n$, we arrive at
\begin{align}
&
\sum_{\alpha_1,\alpha_2,\dots,\alpha_n=\pm}^{\alpha_1\alpha_2\cdots\alpha_n=+}
C_{AB}^{\alpha_1\alpha_2\cdots\alpha_n}(\omega)
\notag
\\
&\qquad\qquad
=
\coth\left(\frac{\beta\hbar\omega}{2n}\right)
\sum_{\alpha_1,\alpha_2,\dots,\alpha_n=\pm}^{\alpha_1\alpha_2\cdots\alpha_n=-}
C_{AB}^{\alpha_1\alpha_2\cdots\alpha_n}(\omega).
\label{FDT n}
\end{align}
Equation (\ref{FDT n}) is the $n$th-order generalization of the out-of-time-order FDT.
%The even-numbered terms in the binomial expansion $(\{A,B\}+[A,B])^n$ appear on the LHS of Eq.~(\ref{FDT n}),
%while the odd-numbered terms appear on the RHS.
%For $n=1$ and $n=2$, Eq.~(\ref{FDT n}) reduces to the FDTs (\ref{FDT}) and (\ref{FDT2}), respectively.
%For $n=3$, the relation (\ref{FDT n}) reads
%\begin{align}
%&
%C_{\{A,B\}^3}(\omega)+3C_{\{A,B\}[A,B]^2}(\omega)
%\notag
%\\
%&=
%\coth\left(\frac{\beta\hbar\omega}{6}\right)
%\left[
%3C_{\{A,B\}^2[A,B]}(\omega)+C_{[A,B]^3}(\omega)
%\right].
%\label{FDT3}
%\end{align}
%The factor $2$ in Eq.~(\ref{FDT2}) and the factors $3$ in Eq.~(\ref{FDT3})
%are the binomial coefficients for $n=2$ and $n=3$, respectively.

The right-hand side of Eq.~(\ref{FDT n}) can be expressed in the form of a canonical correlation,
similarly to Eq.~(\ref{FDT2 Phi}).
We define an $n$-partite canonical OTOC as
\begin{align}
\Phi_{(AB)^n}(t,t')
&\equiv
\int_0^1 d\lambda {\rm Tr}\left[\left(
\hat\rho^{\frac{1-\lambda}{n}}\hat A(t)
\hat\rho^{\frac{\lambda}{n}}\hat B(t')
\right)^n\right].
\end{align}
Following the same calculation as for $n=2$, we can prove (see Appendix \ref{appendix})
\begin{align}
\sum_{\alpha_1,\alpha_2,\dots,\alpha_n=\pm}^{\alpha_1\alpha_2\cdots\alpha_n=+}
C_{AB}^{\alpha_1\alpha_2\cdots\alpha_n}(\omega)
&=
\frac{2^{n-1}}{n}\beta\hbar\omega\coth\left(\frac{\beta\hbar\omega}{2n}\right)
\Phi_{(AB)^n}(\omega).
\label{FDT n Phi}
\end{align}
In this way, we have obtained infinitely many rigorous equalities [(\ref{FDT n}) and (\ref{FDT n Phi})] for OTOCs.
The classical limit of Eq.~(\ref{FDT n}) formally becomes 
\begin{align}
\partial_t \langle\!\langle A(t)^n B(t')^n\rangle\!\rangle
&=
k_BT \langle\!\langle \{A(t)^n,B(t')^n\}_P\rangle\!\rangle,
\label{classical limit n}
\end{align}
which corresponds to that of the conventional FDT (\ref{FDT}) with $A(t)$ and $B(t)$ replaced
by $A(t)^n$ and $B(t)^n$, respectively.

The other generalization is that the FDT holds not only for OTOCs in the form 
of ${\rm Tr}(\hat\rho^{\frac{1}{2}}\hat A(t)\hat B(t')\hat\rho^{\frac{1}{2}}\hat A(t)\hat B(t'))$
but also in the form of
${\rm Tr}(\hat\rho^{\frac{1-\gamma}{2}}\hat A(t)\hat\rho^{\frac{\gamma}{2}}\hat B(t')\hat\rho^{\frac{1-\gamma}{2}}\hat A(t)\hat\rho^{\frac{\gamma}{2}}\hat B(t'))$, i.e., the operator ordering is rearranged.
For usual time-ordered correlators, this type of rearrangement of operator ordering shows up in the context of
the generalized covariance \cite{Petz2002,Gibilisco2009}
defined by
\begin{align}
C_{AB}^f(t,t')
&\equiv
{\rm Tr}\left(\hat A(t)\bm K_{\hat\rho}^f[\hat B(t')]\right),
\label{GC}
\end{align}
where $\bm K_{\hat\rho}^f \equiv f(\bm L_{\hat\rho}\bm R_{\hat\rho}^{-1})\bm R_{\hat\rho}$
is a super-operator, $\bm R_{\hat\rho}$  ($\bm L_{\hat\rho}$) denotes
an operation of multiplying $\hat\rho$ from the right-hand side (left-hand side),
and $f(x)$ is an operator monotone function satisfying $0\leqslant\hat A\leqslant\hat B\Rightarrow f(\hat A)\leqslant f(\hat B)$. 
Equation (\ref{GC}) generalizes the classical covariance for two observables
that do not necessarily commute with $\hat\rho$.
The generalized covariance has played a key role 
in estimation theory involving the quantum Fisher information \cite{Helstrom1967,YuenLax1973,Petz1996}.
%which is a natural generalization of the classical Fisher information
%that measures a sensitivity of probability distributions to a parameter change in the system.
The conventional FDT (\ref{FDT}) has recently been generalized to \cite{ShitaraUeda2015}
\begin{align}
C_{AB}^f(\omega)
&=
\beta\hbar\omega\frac{f(e^{-\beta\hbar\omega})}{1-e^{-\beta\hbar\omega}}\Phi_{AB}(\omega),
\label{FDT GC}
\end{align}
which provides a means to measure the generalized covariance through 
the response function.
The relation (\ref{FDT Phi}) is a special case of Eq.~(\ref{FDT GC})
with $f(x)=\frac{1+x}{2}$.
%Equation (\ref{FDT GC}) provides a way to measure the generalized covariance through 
%the response function.

With the generalized covariance, the $n$-partite OTOC is generalized in the form of
%another type of higher-order OTOCs as
\begin{align}
C^f_{(AB)^n}(t,t')
&\equiv
{\rm Tr}\left( \left[\hat A(t)\bm K_{\hat\rho^{1/n}}^f[\hat B(t')]\right]^n\right).
\label{GC OTOC}
\end{align}
%There is a class of operator monotone functions where generalized out-of-time-order FDT holds.
In particular, if we take $f(x)=x^\gamma$ with $0\leqslant \gamma \leqslant 1$,
then Eq.~(\ref{GC OTOC}) reads
\begin{align}
C^f_{(AB)^n}(t,t')
=
C^\gamma_{(AB)^n}(t,t')
=
{\rm Tr}\left( 
\left[\hat A(t)\hat\rho^{\frac{\gamma}{n}}\hat B(t')\hat\rho^{\frac{1-\gamma}{n}}
\right]^n\right).
\end{align}
Following similar calculations used in deriving Eqs.~\eqref{reciprocal relation n} and \eqref{FDT n},
we can prove (see Appendix \ref{appendix})
\begin{align}
C^\gamma_{(BA)^n}(\omega)
&=
e^{\frac{1-2\gamma}{n}\beta\hbar\omega}
C^\gamma_{(AB)^n}(-\omega),
\label{reciprocal GC}
\end{align}
which reduces to Eq.~\eqref{reciprocal relation n} for $\gamma=0$
and leads to a generalized out-of-time-order FDT,
\begin{widetext}
\begin{align}
\sum_{\alpha_1,\alpha_2,\dots,\alpha_n=\pm}^{\alpha_1\alpha_2\cdots\alpha_n=+}
C_{AB}^{\gamma,\alpha_1\alpha_2\cdots\alpha_n}(\omega)
&=
\coth\left((1-2\gamma)\frac{\beta\hbar\omega}{2n}\right)
\sum_{\alpha_1,\alpha_2,\dots,\alpha_n=\pm}^{\alpha_1\alpha_2\cdots\alpha_n=-}
C_{AB}^{\gamma,\alpha_1\alpha_2\cdots\alpha_n}(\omega)
\label{GC FDT n}
\\
&\quad=
\frac{2^{n-1}}{n}\beta\hbar\omega \coth\left((1-2\gamma)\frac{\beta\hbar\omega}{2n}\right)
\Phi_{(AB)^n}^\gamma(\omega).
\label{GC FDT n Phi}
\end{align}
\end{widetext}
Here we define
\begin{align}
&
C_{AB}^{\gamma,\alpha_1\alpha_2\cdots\alpha_n}(t,t')
\notag
\\
&\equiv
{\rm Tr}\left(
\prod_{i=1}^n \left[
\hat A(t)\hat\rho^{\frac{\gamma}{n}}\hat B(t')\hat\rho^{\frac{1-\gamma}{n}}
+\alpha_i\hat B(t')\hat\rho^{\frac{\gamma}{n}}\hat A(t)\hat\rho^{\frac{1-\gamma}{n}}
\right]
\right),
\\
&
\Phi_{(AB)^n}^\gamma(t,t')
\equiv
\int_{\gamma}^{1-\gamma} d\lambda {\rm Tr}\left[\left(
\hat\rho^{\frac{1-\lambda}{n}}\hat A(t)
\hat\rho^{\frac{\lambda}{n}}\hat B(t')
\right)^n\right].
\label{Phi n gamma t}
\end{align}
The equality (\ref{GC FDT n}) is the most general form of the out-of-time-order FDT derived in this paper,
which includes Eqs.~(\ref{FDT2}) and (\ref{FDT n}) as special cases.
Let us remark that for $f(x)=x^\gamma$ to be operator monotone we need $0\leqslant \gamma \leqslant 1$. 
However, the equalities (\ref{reciprocal GC}), (\ref{GC FDT n}), and (\ref{GC FDT n Phi}) hold for arbitrary $\gamma\in\mathbb R$.

\section{Conclusion}
\label{conclusion}

In conclusion, we have found the generalized fluctuation-dissipation theorem [Eqs.~(\ref{FDT2}) and (\ref{FDT2 Phi})]
for bipartite out-of-time-ordered correlation functions [Eq.~(\ref{bipartite OTOC})].
The theorem describes the general relationship between chaotic behavior in quantum systems and a nonlinear response.
%behavior of higher-order moments of fluctuations. 
The difference between the bipartite and
physical OTOCs is characterized by the Wigner-Yanase skew information
[Eqs.~(\ref{difference1})-(\ref{difference3})], which quantifies the information contents involved
in the corresponding quantum fluctuations. We have further extended the theorem to
$n$-partite OTOCs [Eqs.~(\ref{FDT n}) and (\ref{FDT n Phi})]
and in the form of the generalized covariance [Eqs.~(\ref{GC FDT n}) and (\ref{GC FDT n Phi})].
%We emphasize that the relation (\ref{FDT2}) can be experimentally tested by using 
%the measurement protocol for bipartite OTOCs proposed in Ref.~\cite{Yao2016}.
Our results bring up various interesting open questions
%that are yet to be clarified,
such as the physical meaning of the higher-order out-of-time-order FDTs ($n\geqslant 3$) 
that are expected to be related to higher-order response functions
%the implication to quantum chaos
%the physical meaning of the classical relations [(\ref{classical limit}) and (\ref{classical limit n})],
and the relation to the fluctuation theorem \cite{EvansCohenMorriss1993,EvansSearles1994}
(see also Ref.~\cite{YungerHalpern2017}), which merit further study.

\acknowledgements

N.T. is supported by JSPS KAKENHI Grant No. JP16K17729.
% and JST, PRESTO.
T.S. acknowledges support from Grant-in-Aid for JSPS Fellows (KAKENHI Grant No. JP16J06936) and the Advanced Leading Graduate Course for Photon Science (ALPS) of JSPS.
M.U. acknowledges support by KAKENHI Grant No. JP26287088 and KAKENHI Grant No. JP15H05855.

\begin{widetext}

\appendix

\section{Proof of Eqs.~(\ref{FDT2 Phi}), (\ref{reciprocal relation n}), (\ref{FDT n Phi}), 
(\ref{reciprocal GC}), (\ref{GC FDT n}), and (\ref{GC FDT n Phi})}
\label{appendix}

In this appendix, we prove the equalities (\ref{reciprocal GC}), (\ref{GC FDT n}), and (\ref{GC FDT n Phi}) given in the main text, i.e.,
\begin{align}
C^\gamma_{(BA)^n}(\omega)
&=
e^{\frac{1-2\gamma}{n}\beta\hbar\omega}C^\gamma_{(AB)^n}(-\omega),
\label{reciprocal n gamma}
\end{align}
and
\begin{align}
\sum_{\alpha_1,\alpha_2,\dots,\alpha_n=\pm}^{\alpha_1\alpha_2\cdots\alpha_n=+}
C_{AB}^{\gamma,\alpha_1\alpha_2\cdots\alpha_n}(\omega)
&=
\coth\left((1-2\gamma)\frac{\beta\hbar\omega}{2n}\right)
\sum_{\alpha_1,\alpha_2,\dots,\alpha_n=\pm}^{\alpha_1\alpha_2\cdots\alpha_n=-}
C_{AB}^{\gamma,\alpha_1\alpha_2\cdots\alpha_n}(\omega)
\label{FDT n gamma}
\\
&=
\frac{2^{n-1}}{n}\beta\hbar\omega \coth\left((1-2\gamma)\frac{\beta\hbar\omega}{2n}\right)
\Phi_{(AB)^n}^\gamma(\omega).
\label{FDT n gamma Phi}
\end{align}
with $n=1,2,3,\dots$. 
%The derivation is parallel to that given in Sec. I.
Here $C^\gamma_{(AB)^n}(\omega)$ and $C^\gamma_{(BA)^n}(\omega)$ in Eq.~(\ref{reciprocal n gamma})
are the Fourier transforms of $n$-partite OTOCs
\begin{align}
C^\gamma_{(AB)^n}(t,t')
&=
{\rm Tr}\left( 
\left[\hat A(t)\hat\rho^{\frac{\gamma}{n}}\hat B(t')\hat\rho^{\frac{1-\gamma}{n}}
\right]^n\right),
\\
C^\gamma_{(BA)^n}(t,t')
&=
{\rm Tr}\left( 
\left[\hat B(t)\hat\rho^{\frac{\gamma}{n}}\hat A(t')\hat\rho^{\frac{1-\gamma}{n}}
\right]^n\right).
\end{align}
Equalities (\ref{reciprocal relation n}) and (\ref{FDT n Phi}) in the main text are the special cases of Eqs.~(\ref{reciprocal n gamma})
and (\ref{FDT n gamma Phi}) with $\gamma=0$, and the equality (\ref{FDT2 Phi}) in the main text is the special case of Eq.~(\ref{FDT n gamma Phi})
with $n=2$ and $\gamma=0$.

First, we expand $C^\gamma_{(AB)^n}(t,t')$ and $C^\gamma_{(BA)^n}(t,t')$ in the basis of the eigenstates of the Hamiltonian $\hat H$,
\begin{align}
C^\gamma_{(AB)^n}(t,t')
&=
\frac{1}{Z}\sum_{i_1\cdots i_{2n}}
e^{-\frac{(1-\gamma)\beta}{n}(E_{i_1}+E_{i_3}+\cdots+E_{i_{2n-1}})-\frac{\gamma\beta}{n}(E_{i_2}+E_{i_4}+\cdots+E_{i_{2n}})}
e^{\frac{i}{\hbar}(E_{i_1}-E_{i_2}+E_{i_3}-E_{i_4}+\cdots+E_{i_{2n-1}}-E_{i_{2n}})(t-t')} 
\notag
\\
&\qquad\quad\quad\times
\langle i_1|\hat A|i_2\rangle
\langle i_2|\hat B|i_3\rangle
\langle i_3|\hat A|i_4\rangle
\langle i_4|\hat B|i_5\rangle
\cdots
\langle i_{2n-1}|\hat A|i_{2n}\rangle
\langle i_{2n}|\hat B|i_1\rangle,
\label{C^gamma_(AB)^n}
\\
C^\gamma_{(BA)^n}(t,t')
&=
\frac{1}{Z}\sum_{i_1\cdots i_{2n}}
e^{-\frac{(1-\gamma)\beta}{n}(E_{i_1}+E_{i_3}+\cdots+E_{i_{2n-1}})-\frac{\gamma\beta}{n}(E_{i_2}+E_{i_4}+\cdots+E_{i_{2n}})}
e^{\frac{i}{\hbar}(E_{i_1}-E_{i_2}+E_{i_3}-E_{i_4}+\cdots+E_{i_{2n-1}}-E_{i_{2n}})(t-t')} 
\notag
\\
&\qquad\quad\quad\times
\langle i_1|\hat B|i_2\rangle
\langle i_2|\hat A|i_3\rangle
\langle i_3|\hat B|i_4\rangle
\langle i_4|\hat A|i_5\rangle
\cdots
\langle i_{2n-1}|\hat B|i_{2n}\rangle
\langle i_{2n}|\hat A|i_1\rangle.
\label{C^gamma_(BA)^n}
\end{align}
Then we permute the summation indices as $i_1 \to i_2 \to i_3 \to \cdots \to i_{2n-1} \to i_{2n} \to i_1$
in Eq.~(\ref{C^gamma_(BA)^n}), obtaining
\begin{align}
C_{(BA)^n}^\gamma(t,t')
&=
\frac{1}{Z}\sum_{i_1\cdots i_{2n}}
e^{-\frac{(1-\gamma)\beta}{n}(E_{i_2}+E_{i_4}+\cdots+E_{i_{2n}})-\frac{\gamma\beta}{n}(E_{i_1}+E_{i_3}+\cdots+E_{i_{2n-1}})}
e^{-\frac{i}{\hbar}(E_{i_1}-E_{i_2}+E_{i_3}-E_{i_4}+\cdots+E_{i_{2n-1}}-E_{i_{2n}})(t-t')} 
\notag
\\
&\qquad\quad\quad\times
\langle i_1|\hat A|i_2\rangle
\langle i_2|\hat B|i_3\rangle
\langle i_3|\hat A|i_4\rangle
\langle i_4|\hat B|i_5\rangle
\cdots
\langle i_{2n-1}|\hat A|i_{2n}\rangle
\langle i_{2n}|\hat B|i_1\rangle.
\label{C_(BA)^n 1.5}
\end{align}
Fourier transforming Eqs.~(\ref{C^gamma_(AB)^n}) and (\ref{C_(BA)^n 1.5}), we obtain
\begin{align}
C_{(AB)^n}^\gamma(\omega)
&=
\frac{1}{Z}\sum_{i_1\cdots i_{2n}}
e^{-\frac{(1-\gamma)\beta}{n}(E_{i_1}+E_{i_3}+\cdots+E_{i_{2n-1}})-\frac{\gamma\beta}{n}(E_{i_2}+E_{i_4}+\cdots+E_{i_{2n}})}
2\pi\delta\!\left(\omega+\tfrac{1}{\hbar}(E_{i_1}-E_{i_2}+E_{i_3}-E_{i_4}+\cdots+E_{i_{2n-1}}-E_{i_{2n}})\right)
\notag
\\
&\qquad\quad\quad\times
\langle i_1|\hat A|i_2\rangle
\langle i_2|\hat B|i_3\rangle
\langle i_3|\hat A|i_4\rangle
\langle i_4|\hat B|i_5\rangle
\cdots
\langle i_{2n-1}|\hat A|i_{2n}\rangle
\langle i_{2n}|\hat B|i_1\rangle,
\label{C_(AB)^n 2}
\\
C_{(BA)^n}^\gamma(\omega)
&=
\frac{1}{Z}\sum_{i_1\cdots i_{2n}}
e^{-\frac{(1-\gamma)\beta}{n}(E_{i_2}+E_{i_4}+\cdots+E_{i_{2n}})-\frac{\gamma\beta}{n}(E_{i_1}+E_{i_3}+\cdots+E_{i_{2n-1}})}
2\pi\delta\!\left(\omega-\tfrac{1}{\hbar}(E_{i_1}-E_{i_2}+E_{i_3}-E_{i_4}+\cdots+E_{i_{2n-1}}-E_{i_{2n}})\right)
\notag
\\
&\qquad\quad\quad\times
\langle i_1|\hat A|i_2\rangle
\langle i_2|\hat B|i_3\rangle
\langle i_3|\hat A|i_4\rangle
\langle i_4|\hat B|i_5\rangle
\cdots
\langle i_{2n-1}|\hat A|i_{2n}\rangle
\langle i_{2n}|\hat B|i_1\rangle.
\label{C_(BA)^n 2}
\end{align}
Due to the presence of the $\delta$ function, we can replace $E_{i_2}+E_{i_4}+\cdots+E_{i_{2n}}$
with $E_{i_1}+E_{i_3}+\cdots+E_{i_{2n-1}}-\hbar\omega$,
and
$E_{i_1}+E_{i_3}+\cdots+E_{i_{2n-1}}$
with $E_{i_2}+E_{i_4}+\cdots+E_{i_{2n}}+\hbar\omega$
in the exponential in Eq.~(\ref{C_(BA)^n 2}), which results in
\begin{align}
C_{(BA)^n}^\gamma(\omega)
&=
\frac{1}{Z}\sum_{i_1\cdots i_{2n}}
e^{-\frac{(1-\gamma)\beta}{n}(E_{i_1}+E_{i_3}+\cdots+E_{i_{2n-1}}-\hbar\omega)-\frac{\gamma\beta}{n}(E_{i_2}+E_{i_4}+\cdots+E_{i_{2n}}+\hbar\omega)}
2\pi\delta\left(\omega-\tfrac{1}{\hbar}(E_{i_1}-E_{i_2}+E_{i_3}-E_{i_4}+\cdots+E_{i_{2n-1}}-E_{i_{2n}})\right)
\notag
\\
&\qquad\quad\quad\times
\langle i_1|\hat A|i_2\rangle
\langle i_2|\hat B|i_3\rangle
\langle i_3|\hat A|i_4\rangle
\langle i_4|\hat B|i_5\rangle
\cdots
\langle i_{2n-1}|\hat A|i_{2n}\rangle
\langle i_{2n}|\hat B|i_1\rangle.
\end{align}
By comparing this with Eq.~(\ref{C_(AB)^n 2}), we prove the equality (\ref{reciprocal GC}) [i.e., (\ref{reciprocal n gamma})].

Next, we consider
\begin{align}
\sum_{\alpha_1,\alpha_2,\dots,\alpha_n=\pm}^{\alpha_1\alpha_2\cdots\alpha_n=+(-)}
C_{AB}^{\gamma,\alpha_1\alpha_2\cdots\alpha_n}(\omega)
&=
\int_{-\infty}^{\infty} dt\, e^{i\omega t}
\frac{1}{2}
{\rm Tr}\left(
\left[
\left(
\hat A(t)\hat\rho^{\frac{\gamma}{n}}\hat B(0)\hat\rho^{\frac{1-\gamma}{n}}
+\hat B(0)\hat\rho^{\frac{\gamma}{n}}\hat A(t)\hat\rho^{\frac{1-\gamma}{n}}
\right)
+\left(
\hat A(t)\hat\rho^{\frac{\gamma}{n}}\hat B(0)\hat\rho^{\frac{1-\gamma}{n}}
-\hat B(0)\hat\rho^{\frac{\gamma}{n}}\hat A(t)\hat\rho^{\frac{1-\gamma}{n}}
\right)
\right]^n
\right.
\notag
\\
&\quad\left.
+(-)\left[
\left(
\hat A(t)\hat\rho^{\frac{\gamma}{n}}\hat B(0)\hat\rho^{\frac{1-\gamma}{n}}
+\hat B(0)\hat\rho^{\frac{\gamma}{n}}\hat A(t)\hat\rho^{\frac{1-\gamma}{n}}
\right)
-\left(
\hat A(t)\hat\rho^{\frac{\gamma}{n}}\hat B(0)\hat\rho^{\frac{1-\gamma}{n}}
-\hat B(0)\hat\rho^{\frac{\gamma}{n}}\hat A(t)\hat\rho^{\frac{1-\gamma}{n}}
\right)
\right]^n
\right)
\notag
\\
&=
\int_{-\infty}^{\infty} dt\, e^{i\omega t}2^{n-1}
{\rm Tr}\left(
\left[
\hat A(t)\hat\rho^{\frac{\gamma}{n}}\hat B(0)\hat\rho^{\frac{1-\gamma}{n}}
\right]^n
+(-)\left[
\hat B(0)\hat\rho^{\frac{\gamma}{n}}\hat A(t)\hat\rho^{\frac{1-\gamma}{n}}
\right]^n
\right)
\notag
\\
&=
2^{n-1}\left[
C_{(AB)^n}^\gamma(\omega)+(-)C_{(BA)^n}^\gamma(-\omega)
\right]
\notag
\\
&=
2^{n-1}\left[1+(-)e^{-\frac{(1-2\gamma)}{n}\beta\hbar\omega}\right]
C_{(AB)^n}^\gamma(\omega).
\label{FDT n gamma2}
\end{align}
Here we have used the relation (\ref{reciprocal GC}) in deriving the fourth equality.
Taking the ratio of both sides of Eq.~(\ref{FDT n gamma2})
between the ones with $+$ and $-$ signs proves Eq.~(\ref{GC FDT n}) [i.e., (\ref{FDT n gamma})].

Finally, we expand the partial $n$-partite canonical OTOC $\Phi_{(AB)^n}^\gamma(t,t')$ defined by Eq.~(\ref{Phi n gamma t}) in the basis
of the eigenstates of the Hamiltonian $\hat H$,
\begin{align}
\Phi_{(AB)^n}^\gamma(t,t')
&=
\frac{1}{Z}\int_{\gamma}^{1-\gamma} d\lambda \sum_{i_1\cdots i_{2n}}
e^{-\frac{\beta}{n}(E_{i_1}+E_{i_3}+\cdots+E_{i_{2n-1}})}
e^{\frac{\lambda\beta}{n}(E_{i_1}-E_{i_2}+E_{i_3}-E_{i_4}+\cdots+E_{2n-1}-E_{2n})} 
e^{\frac{i}{\hbar}(E_{i_1}-E_{i_2}+E_{i_3}-E_{i_4}+\cdots+E_{2n-1}-E_{2n})(t-t')}
\notag
\\
&\qquad\qquad\quad\quad\quad\times
\langle i_1|\hat A|i_2\rangle \langle i_2|\hat B|i_3\rangle
\langle i_3|\hat A|i_4\rangle \langle i_4|\hat B|i_5\rangle
\cdots \langle i_{2n-1}|\hat A|i_{2n}\rangle \langle i_{2n}|\hat B|i_1\rangle,
\end{align}
which is Fourier transformed into
\begin{align}
\Phi_{(AB)^n}^\gamma(\omega)
&=
\frac{1}{Z}\int_{\gamma}^{1-\gamma} d\lambda e^{-\frac{\lambda\beta}{n}\hbar\omega} \sum_{i_1\cdots i_{2n}}
e^{-\frac{\beta}{n}(E_{i_1}+E_{i_3}+\cdots+E_{i_{2n-1}})}
2\pi\delta\!\left(\omega+\tfrac{1}{\hbar}(E_{i_1}-E_{i_2}+E_{i_3}-E_{i_4}+\cdots-E_{2n})\right)
\notag
\\
&\qquad\qquad\qquad\quad\quad\quad\quad\times
\langle i_1|\hat A|i_2\rangle \langle i_2|\hat B|i_3\rangle 
\langle i_3|\hat A|i_4\rangle \langle i_4|\hat B|i_5\rangle
\cdots \langle i_{2n-1}|\hat A|i_{2n}\rangle \langle i_{2n}|\hat B|i_1\rangle
\notag
\\
&=
\frac{1}{Z} \frac{e^{-\frac{\gamma}{n}\beta\hbar\omega}-e^{-\frac{(1-\gamma)}{n}\beta\hbar\omega}}{\beta\hbar\omega/n} \sum_{i_1\cdots i_{2n}}
e^{-\frac{\beta}{n}(E_{i_1}+E_{i_3}+\cdots+E_{i_{2n-1}})}
2\pi\delta\!\left(\omega+\tfrac{1}{\hbar}(E_{i_1}-E_{i_2}+E_{i_3}-E_{i_4}+\cdots-E_{2n})\right)
\notag
\\
&\qquad\qquad\qquad\qquad\qquad\times
\langle i_1|\hat A|i_2\rangle \langle i_2|\hat B|i_3\rangle 
\langle i_3|\hat A|i_4\rangle \langle i_4|\hat B|i_5\rangle
\cdots \langle i_{2n-1}|\hat A|i_{2n}\rangle \langle i_{2n}|\hat B|i_1\rangle.
\end{align}
By comparing this with Eq.~(\ref{C_(AB)^n 2}), we obtain
\begin{align}
\Phi_{(AB)^n}^\gamma(\omega)
&=
\frac{n}{\beta\hbar\omega}
\left[1-e^{-\frac{(1-2\gamma)}{n}\beta\hbar\omega}\right]C_{(AB)^n}^\gamma(\omega).
\label{Phi n gamma}
\end{align}
The combination of Eqs.~(\ref{FDT n gamma2}) and (\ref{Phi n gamma})
proves Eq.~(\ref{GC FDT n Phi}) [i.e., (\ref{FDT n gamma Phi})]. 
\hspace{\fill}$\blacksquare$
\end{widetext}

\bibliographystyle{apsrev}
\bibliography{ref}

\end{document}